\documentclass{article}
 \usepackage[final,nonatbib]{neurips_2020}
\usepackage{graphicx,color,xcolor}
\usepackage{hyperref}

\newcommand{\ignore}[1]{}
\usepackage{multirow}
\usepackage{amsfonts}
\usepackage{tcolorbox}
\tcbuselibrary{breakable,xparse,skins}
\definecolor{bg}{gray}{0.95}
\usepackage{color}

\definecolor{bg}{gray}{0.95}
\newcommand{\TW}{{\em Tile-Wise}}
\newcommand{\PW}{{\em Patient-Wise}}
\newcommand{\RL}{RLab}
\newcommand{\MCCv}{$\mathrm{MCC_v}$}
\newcommand{\MCCt}{$\mathrm{MCC_t}$}
\newcommand{\ACCv}{$\mathrm{ACC_v}$}
\newcommand{\ACCt}{$\mathrm{ACC_t}$}
\newcommand{\MCCRL}{$\mathrm{MCC_{RL}}$}
\newcommand{\ACCRL}{$\mathrm{ACC_{RL}}$}

\definecolor{mygreen}{RGB}{0, 128, 0}

\usepackage{caption} 
\begin{document}
\title{AI slipping on tiles:\\ data leakage in digital pathology}
%
%\titlerunning{Abbreviated paper title}
% If the paper title is too long for the running head, you can set
% an abbreviated paper title here
%

% \href{https://orcid.org/0000-0002-2609-4354}{\textcolor{orcidlogocol}{\aiOrcid} \hspace{2mm} orcid.org/0000-0002-2609-4354}

% 

\author{%
Nicole Bussola${}^*\diamond$\\
Fondazione Bruno Kessler, Trento, Italy,\\ University of Trento, Trento, Italy\\
 \texttt{bussola@fbk.eu}\\
\And
Alessia Marcolini${}^*$\\
HK3 Lab, Milano, Italy\\
\texttt{alessia.marcolini@hk3lab.ai}\\
\And
Valerio Maggio\\
University of Bristol, Bristol, United Kingdom\\
\texttt{valerio.maggio@bristol.ac.uk}\\
\And
Giuseppe Jurman${}^\dagger$\\
Fondazione Bruno Kessler, Trento, Italy\\
\texttt{jurman@fbk.eu}\\
\AND
Cesare Furlanello${}^\dagger$\\
HK3 Lab, Milano, Italy\\
\texttt{cesare.furlanello@hk3lab.ai}\\
${}^*$ joint first author, ${}^\dagger$ joint last author, ${}^\diamond$ Corresponding author
}

\maketitle              % typeset the header of the contribution
\begin{abstract}
Reproducibility of AI models on biomedical data still stays as a major concern for their acceptance into the clinical practice. Initiatives for reproducibility in the development of predictive biomarkers as the MAQC Consortium already underlined the importance of appropriate Data Analysis Plans (DAPs) to control for different types of bias, including data leakage from the training to the test set. In the context of digital pathology, the leakage typically lurks in weakly designed experiments not 
accounting for the subjects in their data partitioning schemes. This issue is then exacerbated when fractions or subregions of slides (i.e. ``tiles'') are considered. Despite this aspect is largely recognized by the community, we argue that it is often overlooked. In this study, we assess the impact of data leakage on the performance of machine learning models trained and validated on  multiple histology data collection. We prove that, even with a properly designed DAP ($10 \times 5 $ repeated cross-validation), predictive scores can be inflated up to $41\%$ when tiles from the same subject are used both in training and validation sets by deep learning models. We replicate the experiments for $4$ classification tasks on $3$ histopathological datasets, for a total of $374$ subjects, $556$ slides and more than $27,000$ tiles. Also, we discuss the effects of data leakage on transfer learning strategies with models pre-trained on general-purpose datasets or off-task digital pathology collections. Finally, we propose a solution that automates the creation of leakage-free deep learning pipelines for digital pathology based on {\tt histolab}, a novel Python package for histology data preprocessing. We validate the solution on two public datasets (TCGA and GTEx).

\bf{keywords:} reproducibility, deep learning, digital pathology.
\end{abstract}

\section{Introduction}
Bioinformatics on high-throughput omics data has been plagued by uncountable issues with reproducibility since its early days; 
Ioannidis and colleagues~\cite{ioannidis2009repeatability} found that almost 90\% of papers in a leading journal in genetics were not repeatable due to methodological or clerical errors. Although the landscape seems to have improved~\cite{iqbal2016reproducible}, and broad efforts have been spent across different biomedical fields~\cite{national2019reproducibility}, computational reproducibility and replicability still fall short of the ideal. Lack of reproducibility has been linked to inaccuracies in managing batch 
effects~\cite{leek2010tackling,moossavi2020repeatability}, small sample sizes~\cite{turner2018small}, or flaws in the experimental design such as data normalization simultaneously performed on development and validation data~\cite{barla2008machine,peixoto2015data}. The MAQC-II project for reproducible biomarker development from microarray data demonstrated, through a community-wide research effort, that a well-designed Data Analysis Plan (DAP) is mandatory to avoid selection bias flaws in the development of models for high-dimensional datasets~\cite{maqc10maqcII}.

Among the various types of selection bias that threaten the reproducibility of machine learning algorithms, \emph{data leakage} is possibly the most subtle one~\cite{ching2018opportunities}. Data leakage refers to the use of information from outside the training dataset during model training or selection~\cite{saravanan2018data}. A typical leakage occurs when data in the training, validation and/or test sets share indirect information, leading to overly optimistic results. For example, one of the preclinical sub-dataset in the MAQC-II study consisted of microarray data from mice triplets. These triplets were expected to have an almost identical response for each experimental condition, and therefore they had to be kept together in DAP partitioning to circumvent any possible leakage from training to internal validation data~\cite{maqc10maqcII}.

The goal of this study is to provide evidence that similar issues are still lurking in the grey areas of preprocessing, ready to emerge in the everyday practice of machine learning for digital pathology. The BreaKHis~\cite{spanhol2016dataset} dataset, one of the most popular histology collection of breast cancer samples, has been used in more than $40$ scientific papers to date~\cite{shahidi2020}, with reported results spanning a broad range of performance. In a non-negligible number of these studies, overfitting effects due to data leakage are suspected to impact their outcomes.%~\cite{li2018multitask,nawaz2018multi,xie2019deep,jiang2019breast,jannesary2018,mehra2018breast}.

Deep learning pipelines for histopathological data typically require Whole Slide Images (WSIs) to be partitioned into multiple patches (also referred to as ``tiles''~\cite{cohen2020artificial}) to augment the original training data, and to comply with memory constraints imposed by GPU hardware architectures. For example, a single WSI of size $67,727\times 47,543$ pixels can be partitioned in multiple $512 \times 512$ tiles, which are randomly extracted, and verified such that selected subregions preserve enough tissue information. These tiles are then processed by data augmentation operators (e.g. random rotation, flipping, or affine transformation) to reduce the risk of overfitting. As a result, the number of multiple subimages originating from the very same histological specimen is significantly amplified~\cite{komura2018machine,mormont2018comparison}, consequently increasing the the risk for data leakage. Protocols for data partitioning (e.g. a repeated cross-validation DAP) are not naturally immune against replicates, and so the source originating each tile should be considered to avoid any risk of bias~\cite{maree2017need}. 
% Such bias will inflate the accuracy estimates on the development data, preventing generalization on novel held-out data. 

\begin{figure}[!thb]
    \centering
    \includegraphics[scale=0.15]{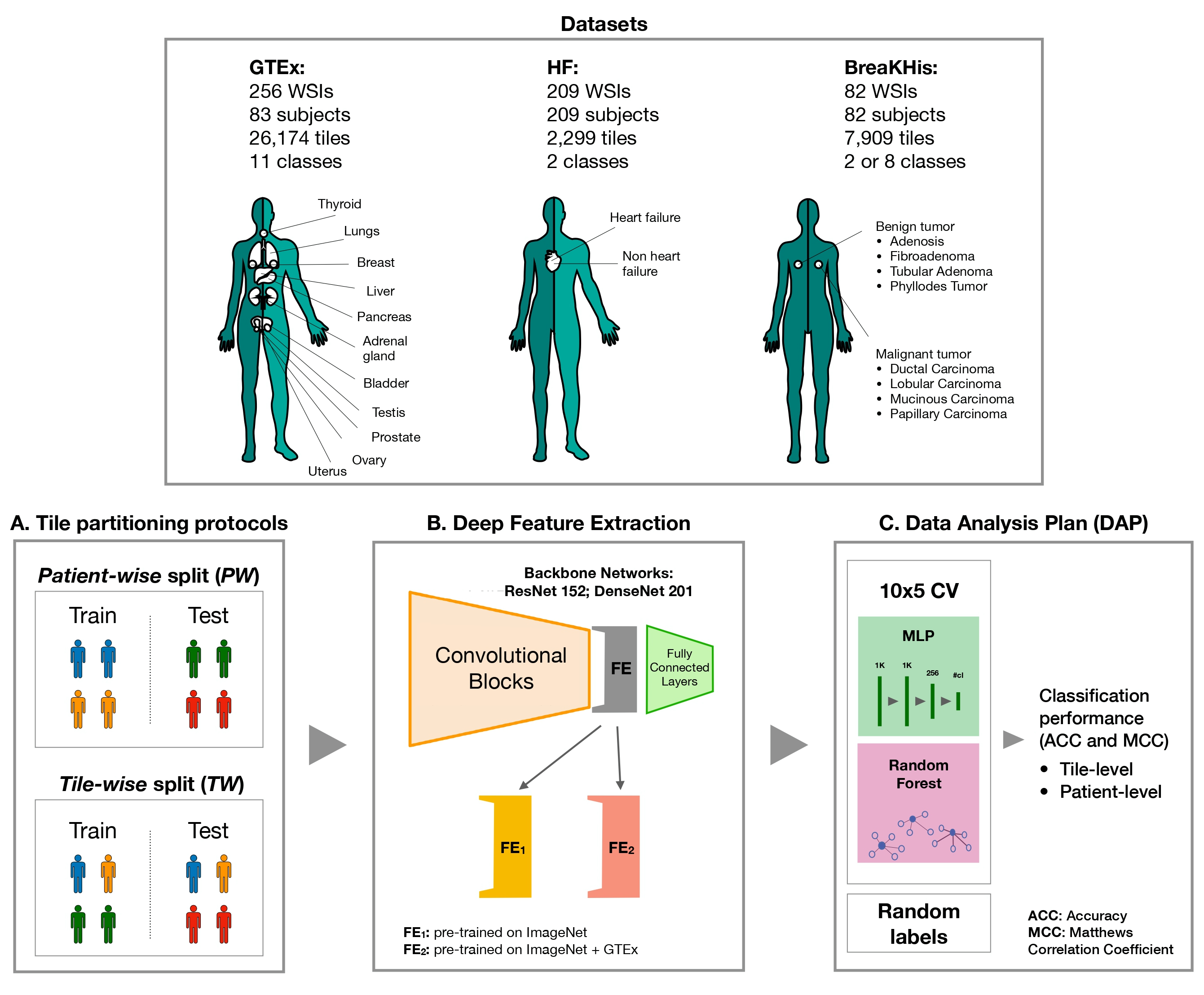}
    \caption{Experimental environment for evaluation of data leakage impact on machine learning models in digital pathology. (A) Tile datasets are split into train/test set following either the \TW\ or the \PW\ protocol; (B) the train set is used to train a backbone network for feature extraction, using different transfer learning strategies; (C) machine learning classifiers on the deep features are evaluated within the Data Analysis Plan.}
    \label{fig:pipeline}
\end{figure}

In this work, we quantify the importance of adopting \PW\ split procedures with a set of experiments on digital pathology datasets. All experiments are based on DAPPER~\cite{bizzego2019evaluating}, a reproducible framework for predictive digital pathology composed of a deep learning core ("backbone network") as feature encoder, and multiple task-related classification models, i.e.~Random Forest or Multi-Layer Perceptron Network (see Fig.~\ref{fig:pipeline}). We test the impact of various data partitioning strategies on the training of multiple backbone architectures, i.e.~DenseNet~\cite{huang2018densely}, and ResNet models~\cite{he2016deep}, fine-tuned to the histology domain. 

Our experiments confirm that train-test  contamination (in terms of modeling) is a serious concern that hinders the development of a dataset-agnostic methodology, with impact similar to the lack of standard protocols in the acquisition and storage of WSIs in digital pathology~\cite{barisoni2020digital}. Thus, we present a protocol to prevent data leakage during data preprocessing. The solution is based on {\tt histolab}, an open-source Python library designed as a reproducible and robust environment for WSI preprocessing, available at \url{https://github.com/histolab/histolab}. The novel approach is demonstrated on two public large scale datasets: GTEx~\cite{lonsdale2013genotype} (i.e.~non-pathological tissues), and TCGA~\cite{tomczak2015cancer} (i.e.~cancer tissues). 
\section{Data description}
We tested our experimental pipeline on three public datasets for image classification in digital pathology, namely GTEx~\cite{lonsdale2013genotype}, Heart Failure (HF)~\cite{nirschl2018deep}, and  
BreaKHis~\cite{spanhol2016dataset}. Descriptive statistics of the datasets are reported in Table~\ref{tab:stats1}, and Fig.\ref{fig:pipeline}.

\begin{table}[!t]
    \centering
    \scriptsize{
	    \begin{tabular}{|c|c|c|c|c|c|c|c|c|c|}
	        \hline
	        \multirow{2}{*}{{\bf Dataset}} & 
	        \multirow{2}{*}{{\bf Subjects}} &
	        \multirow{2}{*}{{\bf WSIs}} & 
	        \multicolumn{3}{|c|}{{\bf WSIs per Subject}} &
	        \multirow{2}{*}{{\bf Tiles}} &
	        \multicolumn{3}{|c|}{{\bf Tiles per Subject}}
	        \\
	        \cline{4-6}\cline{8-10}
	        & & & {Min} & {Max} & {Median} & 
	        & { Min} & { Max} & {Median}
	        \\ \hline
	        GTEx & $83$
	        & $265$ & $1$ & $7$ & $3$ & $26,174$ & $1$ & $700$ & $300$ 
	        \\ \hline
	        HF & $209$ & $209$ 
	        & \multicolumn{3}{|c|}{$1$} & $2,299$ & 
	        \multicolumn{3}{|c|}{$11$} \\
	         \hline
	         BreaKHis & $82$ & $82$ &
	         \multicolumn{3}{|c|}{$1$} & $2,013$ & $9$ & $62$ & $21$ 
	         \\ \hline
	    \end{tabular}
	}
	  \caption{Statistics of the datasets considered in this study.}
    \label{tab:stats1}
\end{table}
\setcounter{footnote}{0}
\paragraph{\bf The GTEx dataset} The current release of GTEx (v8) includes a total of $15,201$ H\&E-stained WSIs, retrieved with an Aperio scanner ($20\times$ native magnification) and gathered from a cohort of $838$ nondiseased donors\footnote{\url{https://gtexportal.org/home/releaseInfoPage}}. In this work, we consider a subset of $265$ WSIs randomly selected from $11$ histological classes, for a total of $83$ subjects. From this subset, we randomly selected a balanced number of WSIs per tissue: adrenal gland ($n=24$); bladder ($n=19$); breast ($n=26$); liver ($n=26$); lung ($n=21$); ovary ($n=26$); pancreas ($n=26$); prostate ($n=24$); testis ($n=26$); thyroid ($n=26$); uterus ($n=21$).

We implemented a data preprocessing pipeline to prepare the tile dataset from the GTEx collection.  
First, the tissue region is automatically detected in each WSI; this process combines the {\em Otsu-threshold} binarization method~\cite{otsu1979threshold} with the dilation and hole-filling morphological operations. A maximum of $100$ tiles of size $512 \times 512$ is then randomly extracted from each slide. To ensure that only high-informative images are used, tiles with tissue area that accounts for less than the 85\% of the whole patch are automatically rejected. At the end of this step, a total of $26,174$ random tiles is extracted from the WSIs, each available at different magnification levels (i.e., $20\times, 10\times, 5\times$). In this paper we limit experiments and discussions to tiles at $5\times$ magnification, with no loss of generality.

\paragraph{\bf The HF dataset} The Heart Failure collection~\cite{nirschl2018deep} originates from $209$ H\&E-stained WSIs of the left ventricular tissue, each corresponding to a single subject. The learning task is to distinguish images of {\em heart failure} ($n=94$) from those of {\em non-heart failure} ($n=115$). Slides in the former class are categorized according to the disease subtype: ischemic cardiomyopathy ($n=51$); idiopathic dilated cardiomyopathy ($n=41$); undocumented ($n=2$).
Subjects with no heart failure are further grouped in: normal cardiovascular function ($n=41$); non-HF and no other pathology ($n=72$); non-HF and other tissue pathology ($n=2$).
WSIs in this dataset have been acquired with an Aperio ScanScope at $20\times$ native magnification, and then downsampled at $5\times$ magnification by authors. From each WSI, $11$ non-overlapping patches of size $250\times 250$ were randomly extracted. The entire collection of $2,299$ tiles is publicly available on the Image Data Resource Repository~\footnote{\url{idr.openmicroscopy.org/}} (IDR number: idr0042). 

\paragraph{\bf The BreaKHis dataset} 
The BreakHis histopathological dataset~\cite{spanhol2016dataset} collects $7,909$ H\&E-stained tiles (size $700\times 460$) of malignant or benign breast tumour biopsies. Tiles correspond to regions of interest manually selected by expert pathologists from a cohort of $82$ patients, and made available at different magnification factors, 
i.e., $40\times$, $100\times$, $200\times$, $400\times$)~\cite{spanhol2016dataset}.
To allow for a more extensive comparison with the state of the art, only the $200\times$ magnification factor will be considered in this paper.
The BreakHis dataset currently contains $4$ histological distinct subtypes of benign, and malignant tumours, respectively: Adenosis ($n=444$); Fibroadenoma ($n=1,014$); Tubular Adenoma ($n=453$); Phyllodes Tumor ($n=569$); Ductal Carcinoma ($n=3,451$); Lobular Carcinoma ($n=626$); Mucinous Carcinoma ($n=792$); Papillary Carcinoma ($n=560$). This dataset is used for two classification tasks: (BreaKHis-2) binary classification of benign and malignant tumour samples; (BreaKHis-8) classification of the $8$ distinct tumour subtypes.

\section{Methods}
The pipeline used in this work is based on the DAPPER framework for digital pathology~\cite{bizzego2019evaluating}, extended by (i) integrating specialised train-test splitting protocols, i.e.~\TW\ and \PW; (ii) extending the feature extractor component with new backbone networks; (iii) applying two transfer learning strategies for feature embedding.
Fig.~\ref{fig:pipeline} shows the three main blocks of the experimental environment defined in this paper: (A) dataset partition in train and test set; (B) feature extraction procedure with different transfer learning strategies; (C) the DAP employed for machine learning models.

% \paragraph{\bf A. Tile generation}
% A data preprocessing pipeline is implemented to prepare the tile dataset from the GTEx WSI collection (Fig.~\ref{fig:pipeline}A). First, the tissue region is automatically detected in each WSI; this process combines the {\em Otsu-threshold} binarization method~\cite{otsu1979threshold} with the dilation and hole-filling morphological operations. A maximum of $100$ tiles of size $512\mathrm{px} \times 512\mathrm{px}$ is then randomly extracted from each slide. To ensure that only high-informative images are used, tiles with tissue area that accounts for less than the 85\% of the whole patch are automatically rejected. At the end of this step, the dataset of tiles is generated.
%
\paragraph{\bf A. Dataset partitioning protocols}
The tile dataset is partitioned in the {\em training} set and {\em test} set, considering $80\%$ and $20\%$ split ratio for the two sets, respectively.
We compare two data partitioning protocols to investigate the impact of a train-test contamination (Fig.~\ref{fig:pipeline}A): in the \TW\ (TW) protocol, tiles are randomly split between the training and the test sets, regardless of the original WSI. The \PW\ (PW) protocol splits the tile dataset strictly ensuring that all tiles extracted from the same subject are found either in the training or the test set. To avoid other sources of leakage due to class imbalance~\cite{raschka2018model}, the two protocols are both combined with stratification of samples over the corresponding classes, and any class imbalance is accounted for by weighting the error on generated predictions.

\paragraph{\bf B. Deep Learning models and feature extraction}\label{rlab} The training set is then used to train a deep neural network for feature extraction (Fig.~\ref{fig:pipeline}B), i.e.~a ``backbone'' network whose aim is to learn a vector representation of the data ({\em features embedding}). 
%Most of deep predictive models in digital pathology are variants of the convolutional neural networks (CNNs). In particular, the ResNet models and its variants are widely adopted for detection or classification tasks in histopathology~\cite{bentaieb2019deep}. 
In this study, we consider two backbone architectures in the residual network (ResNet) family, namely \mbox{ResNet-152}~\cite{he2016deep} and \mbox{DenseNet-201}~\cite{huang2018densely}. Given that the DenseNet model has almost the double of parameters\footnote{
\mbox{DenseNet-201}: $\sim 12$M parameters; ResNet-152: $\sim 6$M parameters.}, and so a higher footprint in computational resources, diagnostic experiments and transfer learning are performed only with the ResNet-152 model. Similarly to~\cite{mormont2018comparison}, and~\cite{bizzego2019evaluating}, we started from off-the-shelf version of the models, pre-trained on ImageNet, and then fine-tuned to the digital pathology domain using transfer learning. Specifically, we trained the whole network for $50$ epochs with a learning rate $\eta = 1e-5$, and Adam optimizer~\cite{kingma2014adam}, in combination with the categorical cross-entropy loss. The $\beta_{1}$ and $\beta_{2}$ parameters of the optimizer are respectively set to $0.9$ and $0.999$, with no regularization. To reduce the risk of overfitting, we use train-time data augmentation, namely random rotation and random flipping of the input tiles.

The impact of adopting a single or double-step transfer learning strategy in combination with the \PW\ partitioning protocol is also investigated in this study. Two sets of features embeddings ($FE$) are generated: $FE_{1}$, backbone model fine-tuned from ImageNet; $FE_{2}$, backbone model sequentially fine-tuned from ImageNet and GTEx. 

\paragraph{\bf C. Classification and Data Analysis Plan (DAP)}
The classification is finally performed on the feature embedding within a DAP for machine learning models (Fig.~\ref{fig:pipeline}C). In this work, we compare the performance of two models: Random Forest (RF) and 
Multi-Layer Perceptron Network (MLP). In particular, we apply the 10$\times$5-fold CV schema proposed by the MAQC-II Consortium~\cite{maqc10maqcII}. In the DAP setting, the input datasets are the two separate training and test sets, as resulted from the $80$-$20$ train-test split protocol. The test set is kept completely unseen to the model, and only used for the final evaluation. The training set further undergoes a 5-fold CV iterated $10$ times with a different random seed, resulting in $50$ separated internal {\em validation} sets. These validation sets are generated adopting the same protocols used in the previous train-test generation, namely \TW\ or \PW.
The overall performance of the model is evaluated across all the iterations, in terms of average Matthews Correlation Coefficient (MCC)~\cite{jurmanMCC12} and Accuracy (ACC), both with $95\%$ Studentized bootstrap confidence intervals (CI). Moreover, results have been reported both at tile-level and at patient-level, in order to assess the ability of machine learning models to generalise on unseen subjects (see section~\ref{results}).

\begin{figure}[!ht]
    \centering
    \includegraphics[scale=0.35]{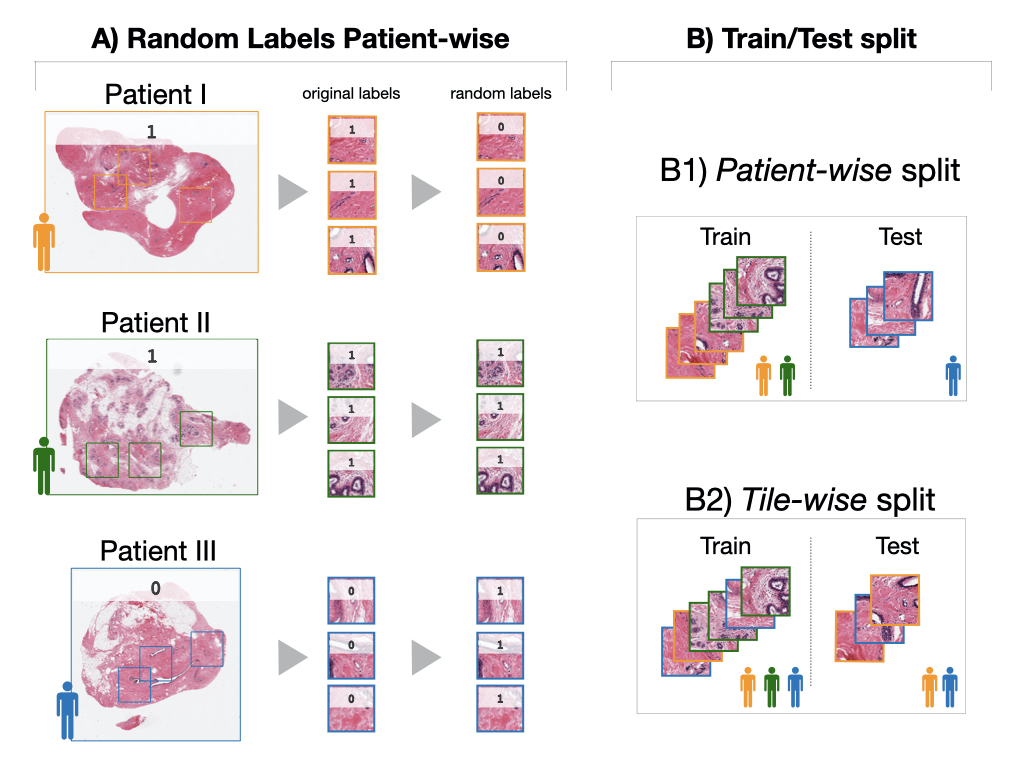}
    \caption{Random Labels experimental settings. A) The labels of the extracted tiles are randomly shuffled consistently with the original patient. B) The train/test split is then performed either {\PW} or {\TW}.}
    \label{fig:RL}
\end{figure}

As an additional caution to check for selection bias, 
the DAP integrates a {\em random labels} schema (\RL) (Fig.~\ref{fig:RL}).
In this setting, the training labels are randomly shuffled and presented as reference ground truth to the machine learning models. In particular, we consistently randomize the labels for all the tiles of a single subject, thus they would all share the same random label (Fig.~\ref{fig:RL}A); then we alternatively use
the \PW\ (Fig.~\ref{fig:RL}B1) or the \TW\ (Fig.~\ref{fig:RL}B2) splits within the DAP environment.
Notice that an average MCC score close to zero ($MCC \approx 0$) indicates a protocol immune from sources of bias, including data leakage; we focus on the \RL\ validation to emphasise evidence of data leakage derived from the {\em TW} and the {\em PW} protocols. 
\paragraph{\bf Performance metrics}
Several patient-wise performance metrics have been defined in the literature~\cite{spanhol2016dataset,alom2019breast,nirschl2018deep}.
Two metrics are considered in this study: 
(1) {\em Winner-takes-all} ($WA$), and (2) {\em Patient Score} ($PS$). 

In the $WA$ metric, the label associated to each patient corresponds to the majority of the labels predicted for their tiles. 
With this strategy, standard metrics based on the classification confusion matrix can be used as overall performance indicators. In this paper, ACC is used for comparability with the $PS$ metric.
The $PS$ metric is defined for each patient~\cite{spanhol2016dataset} as the ratio of the $N_c$ correctly classified tiles over the $N_P$ total number of tiles per patient, namely $PS=\frac{N_c}{N_P}$. The overall performance is then calculated using the {\em global recognition rate} ($RR$), defined as the average of all the $PS$ scores for all patients: 
$$
RR = \displaystyle{\frac{\sum PS}{|P|}}
$$ 
In this paper, the $WA$ metric and the $PS$ metric are used for comparison of patient-level results on the HF dataset and the BreaKHis dataset, respectively.

\paragraph{\bf Preventing Data Leakage: the {\tt histolab} library}
\begin{figure}[!hb]
    \centering
    \includegraphics[scale=0.3]{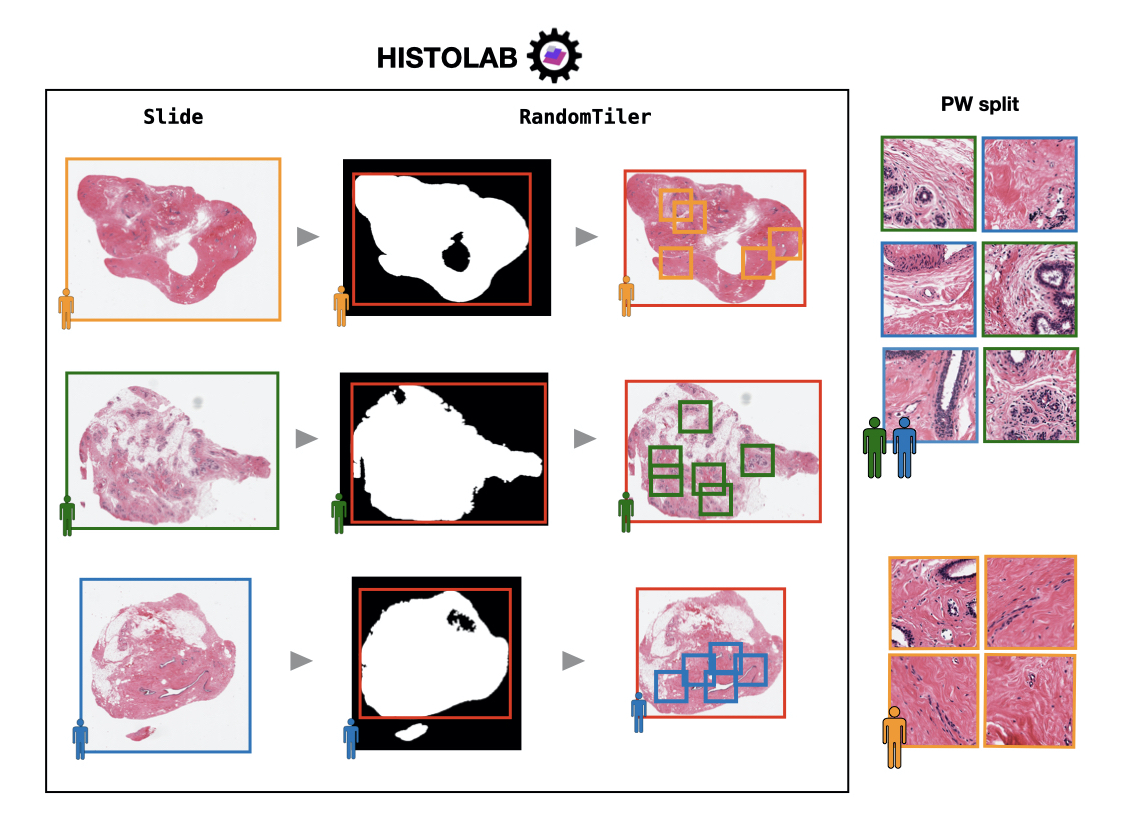}
    \caption{Workflow of the proposed protocol against data leakage in digital pathology, using the histolab software. The documentation of histolab is available at \url{http://histolab.readthedocs.io}.}
    \label{fig:solution}
\end{figure}
As a solution to the data leakage pitfall, we have developed a protocol for image and tile splitting based on 
{\tt histolab}, an open source software recently developed for reproducible WSI preprocessing in digital pathology. This library implements a tile extraction procedure, whose reliability and quality result from robust design, and extensive  software testing. A high level interface for image transformation is also provided, making {\tt histolab} an easy-to-adopt tool for complex histopathological pipelines. 

In order to intercept data leakage conditions, the protocol is designed to create a data-leakage free collection (tile extraction with the \PW\ split) that can be easily integrated in a deep learning workflow (Fig.~\ref{fig:solution}). The protocol is already customized for standardizing WSI preprocessing on GTEx and TCGA, two large scale public repositories that are widely used in computational pathology. The code can be also adapted to rebuild the training and test datasets from GTEx used in this study, thus extending the HINT collection presented in~\cite{bizzego2019evaluating}.

\section{Results}\label{results}
\paragraph{\bf Data Leakage effects on classification outcome}
The results of the four classification tasks using the ResNet-152 pre-trained on ImageNet as backbone model (i.e.~feature vectors $FE_1$) are reported in Table~\ref{tab:results-imagenet-tw} and Table~\ref{tab:results-imagenet-pw}, with the \TW\ and the \PW\ partitioning protocols, respectively. The average cross validation \MCCv\ and \ACCv\ with 95$\%$ CI are presented, along with results on the test set (i.e.~\MCCt, and \ACCt).
State of the art results (i.e.~{\em Others}) are also reported for comparison, whenever available. 
\begin{table}[!t]
\centering
    \tiny{
  \begin{tabular}{|c|cc|cc||cc|cc|c|}
            \cline{1-10} %\hline
            \multirow{2}{*}{\textbf{Dataset}} 
            & \multicolumn{2}{c|}{\textbf{MLP}} 
            & \multicolumn{2}{c||}{\textbf{RF}} 
            & \multicolumn{2}{c|}{\textbf{MLP}} 
            & \multicolumn{2}{c|}{\textbf{RF}} 
            & \textbf{Others} \\
            
            & \MCCv & \MCCt
            & \MCCv & \MCCt
            & \ACCv & \ACCt
            & \ACCv & \ACCt 
            & \ACCt \\ 
            \cline{1-10} %\hline
            
            \multirow{2}{*}{GTEx} 
            & 0.999 & \multirow{2}{*}{0.998}              
            & 0.999 & \multirow{2}{*}{0.997} 
            & 0.999  & \multirow{2}{*}{0.999} 
            & 0.999 & \multirow{2}{*}{0.998} 
            & \multirow{2}{*}{-}  \\
            & (0.999, 0.999) &  
            & (0.999, 0.999) &  
            & (0.999, 0.999) &  
            & (0.999, 0.999) &  
            &  \\ 
            \cline{1-10}
            
            \multirow{2}{*}{HF} 
            & 0.959 & \multirow{2}{*}{0.956} 
            & 0.956 & \multirow{2}{*}{0.960} 
			& 0.980 & \multirow{2}{*}{0.978} 
            & 0.978 & \multirow{2}{*}{0.980} 
            & \multirow{2}{*}{-} \\
            & (0.956, 0.963) &  
			& (0.953, 0.959) &  
			& (0.978, 0.982) &  
			& (0.977, 0.980) &  
			&  \\ 
            \cline{1-10}
             
            \multirow{2}{*}{BreaKHis-2} 
            & 0.989 & \multirow{2}{*}{0.988}  
            & 0.990 & \multirow{2}{*}{0.994}
            & 0.995 & \multirow{2}{*}{0.994}
            & 0.996 & \multirow{2}{*}{0.997} 
            & \multirow{2}{*}{0.993~\cite{jiang2019breast}} \\
            & (0.987, 0.991) &    
            & (0.988, 0.992) &  
            & (0.994, 0.996) &
            & (0.995, 0.997) &  
            &  \\ 
            \cline{1-10}
            
            \multirow{2}{*}{BreaKHis-8} 
            & 0.945 & \multirow{2}{*}{0.922}  
            & 0.929 & \multirow{2}{*}{0.921}
            & 0.959 & \multirow{2}{*}{0.940} 
            & 0.946 & \multirow{2}{*}{0.940} 
            & \multirow{2}{*}{0.985~\cite{jannesary2018}} \\
            & (0.942, 0.949) &    
            & (0.925, 0.932) &
            & (0.956, 0.962) &  
            & (0.943, 0.949) &  
            &  \\ 
            \cline{1-10}
            
%            \hline
        \end{tabular}
    }
    \caption{DAP results for each classifier head, using the \TW\ partitioning protocol, and the $FE_1$ feature embedding with the ResNet-152 as backbone model. The average cross validation metrics (\MCCv\ and \ACCv) with 95$\%$ CI are reported for each classification task, along with metrics on the test set (\MCCt\ and \ACCt). The \textit{Others} column reports the highest accuracy achieved among the compared papers.}
     \label{tab:results-imagenet-tw}
\end{table}

\begin{table}[!thb]
    \centering
    \tiny{
        \begin{tabular}{|c|cc|cc||cc|cc|c|}
            \cline{1-10} %\hline
            \multirow{2}{*}{\textbf{Dataset}} 
            & \multicolumn{2}{c|}{\textbf{MLP}} 
            & \multicolumn{2}{c||}{\textbf{RF}} 
            & \multicolumn{2}{c|}{\textbf{MLP}} 
            & \multicolumn{2}{c|}{\textbf{RF}} 
            & \textbf{Others} 
            \\
            
            & \MCCv & \MCCt
            & \MCCv & \MCCt
            & \ACCv & \ACCt
            & \ACCv & \ACCt 
            & \ACCt \\ 
            \cline{1-10} %\hline
            
            \multirow{2}{*}{GTEx} 
            & 0.998 & \multirow{2}{*}{0.998} 
            & 0.997 & \multirow{2}{*}{0.997} 
            & 0.998 & \multirow{2}{*}{0.998}
            & 0.997 & \multirow{2}{*}{0.997} 
            & \multirow{2}{*}{-}\\
            & (0.998, 0.998) &  
            & (0.997, 0.997) &  
            & (0.998, 0.998) &
            & (0.997, 0.998) &  
            &  \\ 
            \cline{1-10}
             
            \multirow{2}{*}{HF} 
            & 0.852 & \multirow{2}{*}{0.856} 
            & 0.848 & \multirow{2}{*}{0.833} 
            & 0.927 & \multirow{2}{*}{0.915}
            & 0.924 & \multirow{2}{*}{0.915} 
            & \multirow{2}{*}{0.932~\cite{nirschl2018deep}}\\
            & (0.847, 0.858) &  
            & (0.836, 0.860) &  
            & (0.924, 0.929) & 
            & (0.918, 0.930) &  
            &  \\ 
            \cline{1-10}
             
            \multirow{2}{*}{BreaKHis-2} 
            & 0.695 & \multirow{2}{*}{0.801} 
            & 0.709 & \multirow{2}{*}{0.863}
            & 0.870 & \multirow{2}{*}{0.924}  
            & 0.876 & \multirow{2}{*}{0.946} 
            & \multirow{2}{*}{0.973~\cite{alom2019breast}} \\
            & (0.665, 0.724) &    
            & (0.671, 0.746) &
            & (0.856, 0.882) &  
            & (0.859, 0.892) &  
            &  \\ 
            \cline{1-10}
            
            \multirow{2}{*}{BreaKHis-8} 
            & 0.561 & \multirow{2}{*}{0.541}  
            & 0.594 & \multirow{2}{*}{0.471}
            & 0.679 & \multirow{2}{*}{0.644} 
            & 0.701 & \multirow{2}{*}{0.600} 
            & \multirow{2}{*}{0.973~\cite{alom2019breast}} \\
            & (0.529, 0.594) &    
            & (0.562, 0.631) &
            & (0.655, 0.703) &  
            & (0.681, 0.732) &  
            &  \\ 
            \cline{1-10} %\hline
        \end{tabular}
        }
         \caption{DAP results for each classifier head, using the \PW\ partitioning protocol, and the $FE_1$ feature embedding with the ResNet-152 as backbone model. The average cross validation metrics (\MCCv\ and \ACCv) with 95$\%$ CI are reported for each classification task, along with metrics on the test set (\MCCt\ and \ACCt). The \textit{Others} column reports the highest accuracy achieved among the compared papers.}
            \label{tab:results-imagenet-pw}
\end{table}

As expected, estimates are more favourable for the {\em TW} protocol (Tab.~\ref{tab:results-imagenet-tw}) with respect to the {\em PW} one (Tab.~\ref{tab:results-imagenet-pw}), both in validation and in test and consistently for all the datasets.
Moreover, the inflation of the \TW\ estimates is amplified in the 
multi-class setting (see BreaKHis-2 vs BreaKHis-8). 
% , and for both the binary and the multi-class problems. 
Notably, these results are comparable with those in the literature, suggesting the evidence of a data leakage for studies adopting the \TW\ splitting strategy.
Results on the GTEx dataset do not suggest significant differences using the two protocols; however both MCC and ACC metrics lie in a very high range. Analogous results (not reported here) were obtained using the DenseNet-201 backbone model, further confirming the generality of the derived conclusions.
% ~\cite{li2018multitask,alom2019advanced,xie2019deep, jiang2019breast,motlagh2018breast,mehra2018breast}.
% RANDOM LABELS 

\begin{figure}[!hb]
    \centering
    \includegraphics[scale=0.25]{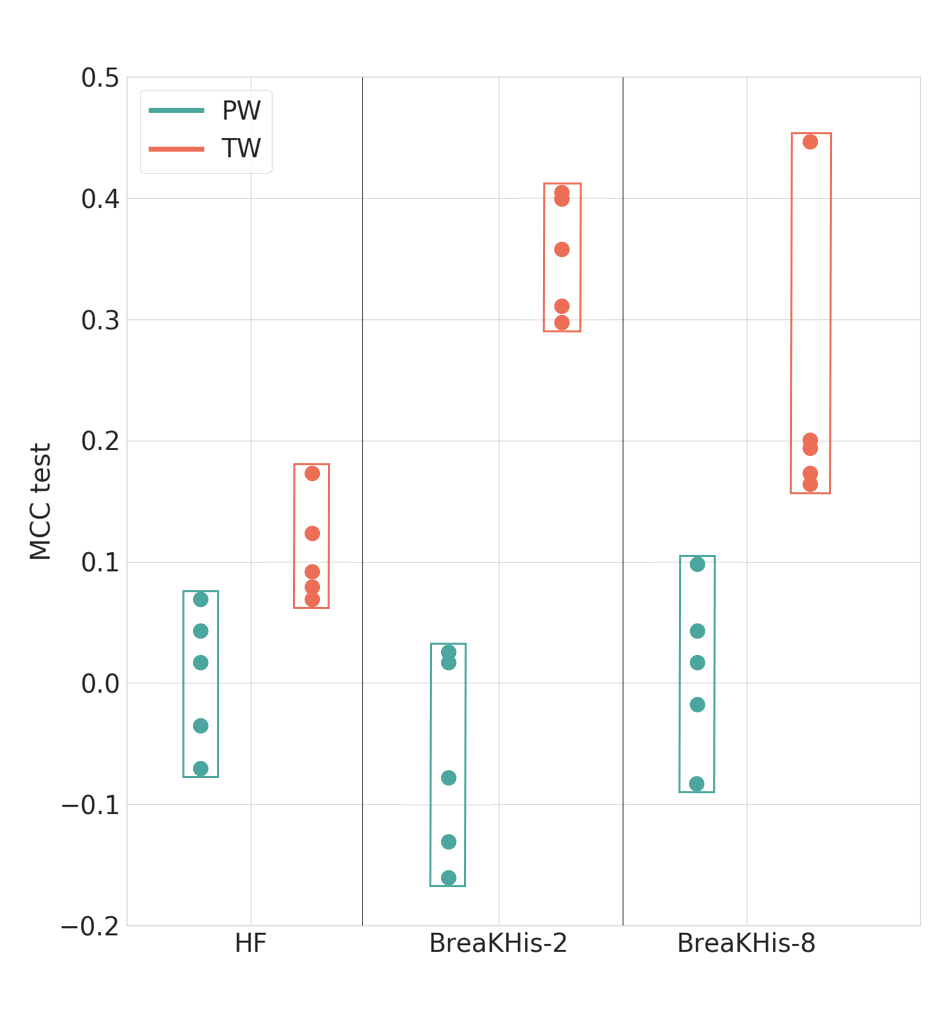}
    \caption{\MCCRL\ results on the test set. TW: \TW, PW: \PW.}
    \label{fig:RL_boxes}
\end{figure}

%
\begin{table}[!t]
	\centering
	\scriptsize{
		\begin{tabular}{|c|cc|cc|}
		\cline{1-5}
		\multirow{2}{*}{\textbf{Dataset}} & 
		\multicolumn{2}{c|}{\textbf{$\mathrm{\mathbf{MCC_{RL}}}$}} & 
		\multicolumn{2}{c|}{\textbf{$\mathrm{\mathbf{ACC_{RL}}}$}} 
		\\
		& {\em TW} & {\em PW}
		& {\em TW} & {\em PW}
		\\
		\cline{1-5}
		
		\multirow{2}{*}{HF}
       	& 0.107 & 0.004
       	& 0.553 & 0.502
       	\\
       	& (0.078, 0.143) & (-0.042, 0.048)
       	& (0.534, 0.570) & (0.474, 0.530)
       	\\
       	\cline{1-5}
		
		\multirow{2}{*}{BreaKHis-2} 
		& 0.354 & -0.065
		& 0.637 & 0.560
		\\
		& (0.319, 0.392) & (-0.131, 0.001)
		& (0.613, 0.662) & (0.506, 0.626)
		\\
		\cline{1-5}
		
		\multirow{2}{*}{BreaKHis-8} 
		& 0.234 & 0.013 
		& 0.318 & 0.097
		\\
		& (0.173, 0.341) & (-0.042, 0.065)
		& (0.215, 0.506) & (0.056, 0.143)
		\\
		\cline{1-5}
		
		\end{tabular}
		}
		\caption{Random Labels (\RL) results using the ResNet-152 as backbone model, 
	         and \TW\ and \PW\ train-test split protocols. 
	         The average \MCCRL\ and \ACCRL\ with 95$\%$ CI are reported.}
	         \label{tab:results-rl}
\end{table}
\paragraph{\bf Random Labels detects signal in the \TW\ split} A data leakage effect is signalled for the \TW\ partitioning with a MCC consistently positive in the \RL\ validation schema (Sect.~\ref{rlab}). 
%
%
%
%\emph{Caveat emptor}: in the experiments with the \RL\ validation schema (see section~\ref{rlab}),  signals are learnt by the model for the \TW\ partitioning, yielding a data leakage effect; in particular, results are consistently overcoming the expected threshold $MCC \approx 0$.
%
For instance, as for BreaKHis-2 coupled with MLP, \MCCRL$=0.354$ $(0.319, 0.392)$ in the \TW\ setting, to be compared with \MCCRL$=-0.065$ $(-0.131, 0.001)$ using the \PW\ protocol. Full \MCCRL\ results considering $5$ trials of the \RL\ test are reported in Table~\ref{tab:results-rl}, with corresponding \ACCRL\ values also included for completeness. Notably, all the tests using the \PW\ split perform as expected, i.e.~with median values near $0$, whereas results of the \TW\ case exhibit a high variability, especially for the BreaKHis-2 dataset (Fig.\ref{fig:RL_boxes}). 
% Transfer Learning FE2

\paragraph{\bf Benefits of domain-specific transfer learning}

\begin{table}[!t]
\centering
\tiny{
    \begin{tabular}{|c|cc|cc||cc|cc|c|}
    	\cline{1-10}
    	\multirow{2}{*}{\textbf{Dataset}} 
        & \multicolumn{2}{c|}{\textbf{MLP}} 
        & \multicolumn{2}{c||}{\textbf{RF}} 
        & \multicolumn{2}{c|}{\textbf{MLP}} 
        & \multicolumn{2}{c|}{\textbf{RF}} 
        & \textbf{Others} 
        \\
        
        & \MCCv & \MCCt
        & \MCCv & \MCCt
        & \ACCv & \ACCt
        & \ACCv & \ACCt 
        & \ACCt \\ 
        \cline{1-10} %\hline
        
        \multirow{2}{*}{HF}  
        & 0.956 & \multirow{2}{*}{0.964} 
        & 0.955 & \multirow{2}{*}{0.950} 
        & 0.978 & \multirow{2}{*}{0.982} 
        & 0.977 & \multirow{2}{*}{0.978} 
        & \multirow{2}{*}{0.932~\cite{nirschl2018deep}} 
        \\   
        & (0.952, 0.960) &  
        & (0.943, 0.958) &  
        & (0.976, 0.980) &  
        & (0.975, 0.979) & 
        & \\
        \cline{1-10}
         
        \multirow{2}{*}{BreaKHis-2} 
        & 0.864 & \multirow{2}{*}{0.948}  
        & 0.912 & \multirow{2}{*}{0.961}
        & 0.941 & \multirow{2}{*}{0.980}
        & 0.963 & \multirow{2}{*}{0.984} 
        & \multirow{2}{*}{0.973~\cite{alom2019breast}} 
        \\  
        & (0.839, 0.888) &  
        & (0.892, 0.932) &  
        & (0.930, 0.952) &
        & (0.955, 0.971) & 
        & \\ 
        \cline{1-10}
        
        \multirow{2}{*}{BreaKHis-8}  
        & 0.573 & \multirow{2}{*}{0.478} 
        & 0.586 & \multirow{2}{*}{0.482}
        & 0.685 & \multirow{2}{*}{0.603}  
        & 0.699 & \multirow{2}{*}{0.606}  
        & \multirow{2}{*}{0.973~\cite{alom2019breast}} 
        \\ 
        & (0.539, 0.602) &  
        & (0.552, 0.621) &  
        & (0.661, 0.712) &  
        & (0.675, 0.724) & 
        & \\ 
        \cline{1-10}
    \end{tabular}
}
 \caption{DAP results for each classifier head, using the \PW\ partitioning protocol, and the $FE_2$ feature embedding with ResNet-152 as backbone model. The average cross validation \MCCv\ and \ACCv\ with 95$\%$ CI are reported, along with results on the test set (i.e.~\MCCt, and \ACCt). The \textit{Others} column reports the highest accuracy achieved among the compared papers.}
  \label{tab:results-fe2}
\end{table}

\begin{table}[!htb]
	\centering
    \scriptsize{
    	\begin{tabular}{|c|c|c|c|c|c|c|c|}
        	\cline{1-8}
        	\multirow{2}{*}{\textbf{Dataset}} & 
        	\textbf{Metric}
        	& \textbf{Partitioning}           
        	& \multicolumn{2}{c|}{\textbf{MLP}}       
        	& \multicolumn{2}{c|}{\textbf{RF}}        
        	& \textbf{Others}
        	\\
        	&                
        	& \textbf{Protocol}
        	& \ACCv           
        	& \ACCt                   
        	& \ACCv           
        	& \ACCt                   
        	& \ACCt                                        
        	\\ 
        	\cline{1-8}
        	\multirow{4}{*}{HF}               
        	& \multirow{4}{*}{$WA$}   
        	& \multirow{2}{*}{TW} 
        	& 0.984          
        	& \multirow{2}{*}{0.995} 
        	& 0.984          
        	& \multirow{2}{*}{0.995} 
        	& \multirow{2}{*}{-}
        	\\
        	
        	& 
        	& & (0.982, 0.987) 
        	& & (0.981, 0.986) 
        	& &
        	\\ 
        	\cline{3-8}
        	                                  
        	& 
        	& \multirow{2}{*}{PW} 
        	& 0.981          
        	& \multirow{2}{*}{0.951} & 0.977          
        	& \multirow{2}{*}{0.927} 
        	& \multirow{2}{*}{0.940~\cite{nirschl2018deep}}
        	\\
        	
        	& 
        	& & (0.975, 0.986) 
        	& & (0.971, 0.983) 
        	& & \\ 
        	\cline{1-8}
        	
        	\multirow{4}{*}{BreaKHis-2}       
        	& \multirow{4}{*}{$PS$}   
        	& \multirow{2}{*}{TW} & 0.995          
        	& \multirow{2}{*}{0.997} & 0.997          
        	& \multirow{2}{*}{0.998} 
        	& \multirow{2}{*}{0.872~\cite{xie2019deep}}                                             
        	\\
        	                                  
            &
            & & (0.994, 0.996) 
            & & (0.996, 0.998) 
            & &
            \\ 
            \cline{3-8}
            & & \multirow{2}{*}{PW} & 0.864          
            & \multirow{2}{*}{0.885} & 0.883
            & \multirow{2}{*}{0.893} & \multirow{2}{*}{0.976~\cite{alom2019breast}}   
            \\
        	
        	&  &
        	& (0.851, 0.877) &     
        	& (0.869, 0.898) & 
        	& \\ 
        	\cline{1-8}
        	
        	\multirow{4}{*}{BreaKHis-8}       
        	& \multirow{4}{*}{$PS$}   & \multirow{2}{*}{TW} & 0.963          
        	& \multirow{2}{*}{0.950} & 0.957          
        	& \multirow{2}{*}{0.962} & \multirow{2}{*}{0.964~\cite{nawaz2018multi}}                           
        	\\
        	& & & (0.960, 0.967) & 
        	& (0.955, 0.959) & &
        	\\ 
        	\cline{3-8}
        	
        	 & & \multirow{2}{*}{PW} & 0.687          
        	 & \multirow{2}{*}{0.752} & 0.705
        	 & \multirow{2}{*}{0.725} & \multirow{2}{*}{0.967~\cite{alom2019breast}}   
        	 \\
        	                                  
        	 &  &  & (0.667, 0.709) & 
        	 & (0.685, 0.728) && 
        	 \\ 
    	    \cline{1-8}
    	\end{tabular}
	}
	\caption{Patient-level results for each classifier head, 
	         using the \PW\ and \TW\ partitioning protocols, 
	         and the $FE_1$ feature embedding with the ResNet-152 backbone model.
             The average cross-validation Patient-level accuracy with 95$\%$ CI (\ACCv), and corresponding scores on the test set (\ACCt), are reported. The \textit{Others} column reports the highest accuracy achieved among the compared papers.}
             \label{tab:resultsPWImageNet}
\end{table}
The adoption of the GTEx domain-specific dataset for transfer learning (Table~\ref{tab:results-fe2}) proves to be beneficial over the use of ImageNet only (Table~\ref{tab:results-imagenet-pw}).
Notably, the \PW\ partitioning protocol with the $FE_{2}$ embedding have comparable performance with $FE_{1}$ and the inflated TW splitting (Tab.~\ref{tab:results-imagenet-tw}).
However, minor improvements are achieved on the BreaKHis-8
task, with results not reaching state of the art. It must be observed that the BreaKHis dataset is highly imbalanced in the multi-class task. 
As a countermeasure, authors in~\cite{han2019breast,alom2019advanced} adopted a balancing strategy during data augmentation, which we did not introduce here for comparability with the other experiments.

To verify how much of previous domain-knowledge can be still re-used for the original task, we devised an additional experiment on the GTEx dataset: on the {\em Feature Extractor} component (i.e.~Convolutional Layers) of the model trained on GTEx and fine-tuned on BreakHis-2, we add back the MLP classifier of the model trained on GTEx. Notably, this configuration recover high predictive performance
(i.e.~\MCCt=0.983) on the classification task after only a single epoch of full training on GTEx.
%Patient Score
\paragraph{\bf Patient-level Performance Analysis}
We report patient-wise performance using the ResNet-152 backbone model with either the $FE_1$ feature embedding and both \TW\ and \PW\ protocols (Table~\ref{tab:resultsPWImageNet}), or with the $FE_2$ strategy and the \PW\ split (Table~\ref{tab:resultsPWGTEx}).
\begin{table}[!ht]
    \centering
\scriptsize{
    \begin{tabular}{|c|c|c|c||c|c|c|}
    \cline{1-7}
    \multirow{2}{*}{\textbf{Dataset}} 
    & \textbf{Patient-level} 
    & \multicolumn{2}{c||}{\textbf{MLP}} 
    & \multicolumn{2}{c|}{\textbf{RF}} & \textbf{Others} 
    \\ 
    & \textbf{Metric} & \ACCv & \ACCt & \ACCv & \ACCt & \ACCv \\ 
      \cline{1-7}
    \multirow{2}{*}{HF} & \multirow{2}{*}{$WA$} & 0.992 & \multirow{2}{*}{0.976} & 0.989 & \multirow{2}{*}{0.976} & \multirow{2}{*}{0.940~\cite{nirschl2018deep}} \\ 
     & & (0.989, 0.995) &  &  (0.984, 0.992) &  &  \\ 
     \cline{1-7}
    \multirow{2}{*}{BreaKHis-2} & \multirow{2}{*}{$PS$} & 0.941 & \multirow{2}{*}{0.971} & 0.958 & \multirow{2}{*}{0.991} & \multirow{2}{*}{0.976~\cite{alom2019breast}} \\ 
    & & (0.930, 0.951) &  & (0.948, 0.968) &  &  \\ 
    \cline{1-7}
    \multirow{2}{*}{BreaKHis-8} & \multirow{2}{*}{$PS$} & 0.691 & \multirow{2}{*}{0.721} & 0.699 & \multirow{2}{*}{0.724} & \multirow{2}{*}{0.967~\cite{alom2019breast}} \\ 
    & & (0.669, 0.716) &  & (0.676, 0.723) &  &  \\ 
    \cline{1-7}
    \end{tabular}
    }
        \caption{Patient-level results for each classifier head, with the \PW\ partitioning protocol and the $FE_2$ feature embedding with the ResNet-152 model. The average cross-validation Patient-level accuracy with 95$\%$ CI (\ACCv) and corresponding scores on the test set (\ACCt) are reported. The \textit{Others} column reports the highest accuracy achieved among the compared papers.}
         \label{tab:resultsPWGTEx}
\end{table}

\section{Discussion}
We report here a short description of the approach employed by comparable studies on the same datasets considered in this work; we refer to a \PW\ partitioning protocol when the authors clearly state the adoption of a train-test split consistent with the patient, or when the code is provided as reference. Notice that the different accuracy scores obtained for deep learning models applied on the same data can be explained by the adoption of diverse experimental protocols (e.g.~preprocessing, data augmentation, transfer learning methods).

Nirschl~et~al.~\cite{nirschl2018deep} train a CNN on the HF dataset to distinguish patients with or without heart failure. They systematically apply the \PW\ rule for the initial train-test split ($50$-$50$) and for the training partition into three-folds for cross-validation. Data augmentation strategies are also applied, including random cropping, rotation, mirroring, and staining augmentation. As for the BreaKHis dataset, Alom~et~al.~\cite{alom2019breast} use a $70$-$30$ \PW\ partitioning protocol to train a CNN with several (not specified) hidden layers, reporting average results from 5-fold cross-validation. Further, the authors apply augmentation strategies (i.e.,~rotation, shifting, flipping) to increase the dataset by a factor of $21\times$ for each magnification level. The work of Han~et~al.~\cite{han2019breast} propose a novel CNN adopting a \TW\ partition with the training set accounting for the 50\% of the dataset. Data augmentation (i.e.~intensity variation, rotation, translation, and flipping) is used to adjust for imbalanced classes. Jiang~et~al.~\cite{jiang2019breast} train two different variants of the ResNet model to address the binary and the multi-class task, for each magnification factor. They adopt a \TW\ partitioning protocol for the train-test split, using 60\% and 70\% of the data in the training set for BreaKHis-2 and BreaKHis-8, respectively. Data augmentation is also exploited in the training process, and experiments are repeated $3$ times.

Other authors employed a similar protocol to address the BreaKHis-8 task by training a CNN pretrained on ImageNet: Nawaz~et~al.~\cite{nawaz2018multi}  implemented a DenseNet-inspired model, while Nguyen~et~al.~\cite{Nguyen2019} choose a custom CNN model, instead. Both studies use a \TW\ partition on the BreaKHis dataset ($70$-$30$ and $90$-$10$, respectively), and do not apply any data augmentation. Xie~et~al.~\cite{xie2019deep} adapt a pre-trained ResNet-V2 to the binary and multiclass tasks of BreaKHis, at different magnification factors, using a $70$-$30$ \TW\ partition. Data augmentation has been applied to balance the least represented class in BreaKHis-8. Jannesary et al.~\cite{jannesary2018} used a $90$-$10$ \TW\ train-test split with data augmentation (i.e.~resizing, rotations, cropping and flipping) to fine-tune a ResNet-V1 for binary and multi-class prediction. Moreover, experiments in~\cite{jannesary2018} were performed combining images at different magnification factors in a unified dataset. Finally, both~\cite{deniz2018transfer} and~\cite{jae2018deep} used a \TW\ train-test split for prediction of malignant vs benign samples using a pre-trained CNN and~\cite{jae2018deep} also employed data augmentation (rotation and flipping).

\section{Conclusions}
Possibly even more than other areas of computational biology, digital pathology faces the risk of data leakage. The first part of this study clearly demonstrates the impact of weakly designed experiments with deep learning for digital pathology. In particular, we found that the predictive performance estimates are inflated if the DAP does not flawlessly concentrate the subject and/or the tissue specimen from which tiles are extracted either in the training or test datasets. Fortunately, many studies already adopt the correct procedure~\cite{mormont2018comparison,maree2017need,spanhol2016dataset,nirschl2018deep,alom2019advanced,han2019breast}. However, we argue that this subtle form of selection bias still constitutes a threat to reproducibility of AI models that may have affected a considerable number of works. Indeed, a significant number of studies considered in this work do not explicitly mention the patient-wise strategy~\cite{li2018multitask,nawaz2018multi,xie2019deep,jiang2019breast,jannesary2018,mehra2018breast}. We encourage the community to adopt our code (\url{https://github.com/histolab/histolab/tree/master/examples}) as a launchpad for reproducibility of AI pipelines in digital pathology.
\bibliographystyle{unsrt}
\bibliography{bussola_slipping}
\end{document}